\documentclass[15pt a4paper]{article}

\usepackage{latexsym}
\usepackage{amsmath,amssymb,amsfonts,amsthm}

\usepackage{graphicx}
\usepackage{subfigure}

\graphicspath{{./}{./figs/}}

\def\marhes{{\sc Marhes~}}
\def\precise{{\sc Precise~}}

\newcommand{\comment}[1]{}
\newcommand{\ie}{{\it i.e.,~}}
\newcommand{\eg}{{\it e.g.,~}}

\renewcommand{\Re}{\mathbb{R}}

\providecommand{\norm}[1]{\left\lVert#1\right\rVert}

\usepackage[T1]{fontenc}
\usepackage[utf8]{inputenc}
\usepackage{authblk}

\begin{document}

\title{Decentralized identification and control of networks of coupled mobile platforms through adaptive synchronization of chaos}


%

\author{Nicola Bezzo$^{\dag}$, Patricio J. Cruz Davalos$^{\ddag}$, Francesco Sorrentino$^{\S}$\thanks{Corresponding Author}, and Rafael Fierro$^{\ddag}$\\
$^{\dag}$\precise Center, Department of Computer and Information Science, \\University of Pennsylvania, Philadelphia, PA 19104, USA\\ {\tt nicbezzo@seas.upenn.edu}\\
$^{\ddag}$\marhes Lab, Department of Electrical and Computer Engineering, \\University of New Mexico, Albuquerque, NM 87131, USA\\ {\tt \{pcruzec, rfierro\}@ece.unm.edu}\\
$^{\S}$Department of Mechanical Engineering, \\University of New Mexico, Albuquerque, NM 87131, USA
        \\{\tt fsorrent@unm.edu}}%

\maketitle
\thispagestyle{empty}
\pagestyle{empty}

\begin{abstract}

In this paper we propose an application of adaptive synchronization of chaos to detect changes in the topology of a mobile robotic network. We assume that the network may evolve in time due to the relative motion of the mobile robots and due to unknown environmental conditions, such as the presence of obstacles in the environment. We consider that each robotic agent is equipped with a chaotic oscillator whose state is propagated to the other robots through wireless communication, with the goal of synchronizing the oscillators.  We introduce an adaptive strategy that each agent independently implements to: (i) estimate the net coupling of all the oscillators in its neighborhood and (ii) synchronize the state of the oscillators onto the same time evolution. We show that by using this strategy, synchronization can be attained and changes in the network topology can be detected. We go one step forward and consider the possibility of using this information to control the mobile network. We show the potential applicability of our technique to the problem of maintaining a formation between a set of mobile platforms, which operate in an inhomogeneous and uncertain environment. We discuss the importance of using chaotic oscillators and validate our methodology by numerical simulations.

{\textit{Keywords:}}  Complex Network, Synchronization. {\bf
89.75.-k,05.45Xt}
\end{abstract}

\section{Introduction} \label{sec:intro}


In the last decade the coordination of networked multi-agent systems has been intensively investigated. Both the robotic and communication research communities have been working on how to properly integrate wireless communication in motion planning algorithms, considering the random properties of the rf channels and the mobility of the autonomous agents.
A typical goal of the research in this area is to accomplish a certain mission while maintaining connectivity of the network. In any real application, the network topology
may change in time due to the random nature of the communication channel and the complexity of the
environment. For example, an agent may remain isolated and become unable to accomplish the mission or
unwanted disconnections may occur due to the presence of obstacles. Moreover, maintaining connectivity
becomes even more challenging when the network architecture is decentralized, for example when each agent has
access only to local information about its connections and the surrounding environment.

Advances in sensor technology in the last three decades have helped
accelerate
interest in the field of robotics. As robots become smaller, more capable,
and less expensive, there is a growing
demand for teams of robots in various application domains. Multi-agent
robotic systems are particularly well suited
to execute tasks that cover wide geographic ranges and/or depend
on capabilities that are varied in both quantity and difficulty. {Example
applications include exploration
and surveillance \cite{burgard2005coordinated}, health monitoring \cite{zhao2008designing}, autonomous transportation systems \cite{ota2006multi},
nanoassembly \cite{galstyan2005modeling} (\ie the programming and coordination of  a large number of
nanorobots), and hazardous waste clean-up \cite{parker1995design}. More recently, research in
multi-agent robotic systems has focused  on robot swarms. Robot swarms are
composed of large numbers of robots capable of covering wide regions and
performing tasks that require significant parallelization and robustness like
parts inspection \cite{baker1998survey}, warehouse automation \cite{monostori2006agent}, and
environmental monitoring \cite{lee2009bio}.}

\begin{figure}[ht!]
\centering
\vspace{6pt}
\includegraphics[width=0.48 \textwidth]{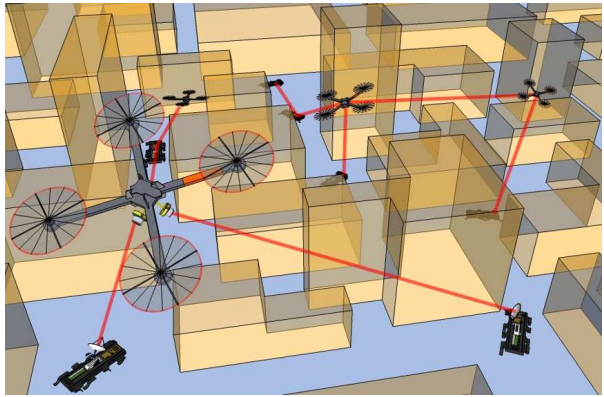}
\caption{\label{fig:env} A heterogeneous network of coupled robotic platforms operating in a cluttered virtual
environment. The lines between the robots represent line-of-sight communication paths. }
\end{figure}

Figure \ref{fig:env} shows a pictorial representation of the situation envisioned in this paper, in which a heterogeneous robotic network operating in a cluttered environment maintains line of sight communication between aerial and ground robots, by using light pulses (\eg lasers transmitters and receivers \cite{rust2012dual,tsai2013los}). Note that communication is selective: only those robots whose line of sight communication paths are not crossed by obstacles are coupled. Also, the network is time-varying as connections can be dynamically created or disrupted as the agents move in different areas of the environment.
 Our underlying assumptions are the following: (i) the robots move in an unknown environment while trying to accomplish a coordinated mission, (ii) they operate in a context characterized by limited availability of information, and (iii) they have no {\em a priori} knowledge of the topology of the network and need to estimate it throughout the mission.

Previous work \cite{SOTT,SOTT2,EXP,EXP2} has shown the possibility of using adaptive synchronization of chaotic systems in a sensor network application, for which the sensors are static. In this paper, we extend this approach to the case that the oscillators are placed on {board} of mobile platforms that move in an unknown, inhomogeneous, and time-varying environment. For this case, we consider that the strengths of the couplings may be affected by the relative motion of the platforms, as well as by the influence of external factors, such as the presence of obstacles in the environment. We will show that our proposed strategy can be employed to detect changes in the network connectivity.

We propose that each mobile robot is equipped with a chaotic oscillator whose state is propagated to the others by wireless communication. At each time, the signal received by each platform is the aggregate chaotic signal received by all the oscillators in its range of communication. Given this available information, we devise a decentralized adaptive strategy that each agent independently implements in order to reach and maintain synchronization while estimating changes in the local connectivity of the network, such as deletion and aggregation of connections.

Moreover, we question how such information can be used to control the network. For example, in many practical applications, an important  goal is that of maintaining the network connectivity, while avoiding collisions between the platforms \cite{ren2008distributed,chen2010leader}. In what follows, we show that a novel chaos-synchronization-based formation control algorithm can be effective in maintaining desired distance and bearing coordinates between a set of mobile robots.


\subsection{Related Work}\label{sec:Relatedwork}

Multiple mobile robotic systems  and wireless communications have been individually and extensively studied for several years. Recently, roboticists have recognized the need to consider realistic communication models when designing multi-robot systems \cite{bezzo2011disjunctive, ghaffarkhah2011communication, hollinger2011autonomous, lindhe2010adaptive, rus2011}.
For instance, the authors of \cite{ ghaffarkhah2011communication} formally analyze the properties of the communication channel and use them to optimally navigate autonomous agents to improve the communication performance in terms of  Signal-to-Noise Ratio and  Bit Error Rate.
In \cite{hollinger2011autonomous} the authors propose a modified Traveling Salesperson Problem to navigate an underwater vehicle in a sensor field, using a realistic model that considers acoustic communication fading effects.
In \cite{rus2011} the researchers present a set of algorithms to repair connectivity within a network of mobile routers and then show outdoor experimental results to validate their algorithms, while in \cite{magnus2011acc} an agent attempts to estimate the network topology by using only local information.

In \cite{bezzo2011disjunctive} a chain of mobile routers is used to keep line of sight communication between a base station and a user  that moves in a concave environment.
In \cite{zavlanosdistributed} the authors optimize routing probabilities to ensure desired communication rates while using a distributed hybrid approach.
In \cite{moore2011} the authors analyze connectivity in consensus networks with dynamical links, while reference \cite{rus2011} presents a set of algorithms to repair connectivity within a network of mobile routers.

Similar to the work presented in this paper, the author in \cite{yu2010estimating} investigate the problem of how to estimate the connection topology in complex dynamical networks by means of a steady-state control based identification method. In \cite{barooah2012cut} the authors propose an algorithm inspired by the Kirchhoff's laws in electrical circuits, to detect losses (also called ``cuts'') in the connectivity of large sensor networks. The same problem is also investigated in \cite{shrivastava2008detecting}, in which a subset of sensors called ``sentinels", communicate on a regular basis to a base station. The authors show that the base station is able to estimate the network topology based only on detections on communication failures with the sentinel nodes.

In a different context, a large literature has investigated synchronization of  networks of coupled dynamical systems \cite{Replace, Pe:Ca, Ba:Pe02, HYP}.
References \cite{SOTT,SAS} present an adaptive technique to synchronize a large sensor network in which aggregate signals are received by each static node of a certain system. This strategy has been  tested in an experimental network of coupled opto-electronic systems in \cite{EXP,EXP2}. A different strategy, aimed at estimating communication-delays, was proposed in \cite{sorrentino2011estimation}.

In our work we consider an extension of the strategy in \cite{SOTT,SAS} and we show that a similar approach can be used to detect changes in networks of coupled mobile robotic platforms.
In practical applications, it is often the case that the robots operate in an unknown cluttered environment and in a decentralized fashion; moreover, the connectivity may be unavailable. In what follows, we will address such situations by introducing a decentralized adaptive technique, which will be proven useful to estimate and track the time-evolution of the network topology.

The remainder of this paper is organized as follows. In Section \ref{sec:Pre} we give some preliminary graph theoretical definitions that will be used throughout the paper. In Section \ref{sec:Sys} we introduce the model for the robot dynamics, the communication connectivity strategy, and the sensing strategy based on the chaotic oscillators. In Sections \ref{sec:AS} we present the adaptive strategy that incorporates the chaotic oscillators, followed by extensive simulation results in Section \ref{sec:Sim}. In Section 6 we discuss the usefulness of dealing with chaotic oscillators. In Section 7 we present a novel chaotic synchronization based formation control algorithm. Finally, conclusions and future work are discussed in Section \ref{sec:Con}.

\section{Preliminaries}\label{sec:Pre}

In this section we introduce the graph theoretical tools \cite{cortezprioritized, graph_theory} to be used in the formulation of the model dynamics and of the decentralized adaptive strategy presented in Section \ref{sec:Sys}.
We proceed under the assumption of planar dynamics. Let ${\mathbf{p}}_i(t)=(y_i(t), z_i(t))^T$ denote the position of robot $i$, $i=1\ldots \mathcal N$ in the $(y, z)$ plane. The network of $\mathcal N$ robots can be represented as a \emph{dynamic graph} $\mathcal{G}(\mathbf{p})$, where ${\mathbf{p}}(t)=({{\mathbf{p}}_1(t)}^T,{{\mathbf{p}}_2(t)}^T,...,{\mathbf{p}_\mathcal{N}(t)}^T)^T$. Namely, we define  $\mathcal{G}(\mathbf{p}) =
(\mathcal{V},\mathcal{E}(\mathbf{p}))$ a dynamic graph consisting of
\begin{itemize}
\item a set of ${\mathcal N}$ vertices $\mathcal{V} = \{v_1,\cdots,v_{\mathcal N}\}$, where each vertex represents a robot, and
\item a set of edges $(v_i,v_j), v_j \neq v_i \in \mathcal{E}(\mathbf{p})$.
\end{itemize}
In our work we assume line-of-sight communication, resulting in undirected edges, though all of our results can be generalized to the case of directed connections.
The existence of an edge $(v_i,v_j)$ depends on the presence or absence of obstacles along the line that connects vertices $i$ and $j$. That is,  $(v_i,v_j) \in \mathcal{E}(\mathbf{p})$ if the line-of-sight communication path that connects $i$ and $j$ is not crossed by any obstacle. A non-empty graph $\mathcal{G}$ is called
\emph{connected} if any two of its vertices are linked by a path in
$\mathcal{G}$.

Given a non-empty dynamic graph $\mathcal{G}$ with
vertices $\mathcal{V}$ and edges in the
set $\mathcal{E}$, we define the adjacency matrix $A(t) = \{A_{ij}(t)\}$
such that $A_{ij}(t) >0$ if $(v_i,v_j) \in \mathcal{E}$, and $A_{ij}(t)
= 0$ otherwise. Moreover, each $A_{ij}(t)$ reflects the strength of the signal between  a receiver node $i$ and a transmitter node $j\neq i$ at time $t$.
 Hence, if $(v_i,v_j) \in \mathcal{E}$, $A_{ij}(t)=P_{ij}(t)$ where $P_{ij}(t)$ is the power received at agent $i$ from transmitter $j$ at time $t$ and $P_{ij}(t)$ is a function of the Euclidean distance $d_{ij}(t)=\norm{\mathbf{p}_i(t) - \mathbf{p}_j(t)}_2$ between platforms $i$ and $j \neq i$. 
 A specific description of how $A_{ij}(t)$ depends on the distance $d_{ij}(t)$ 
 is provided in Sec. \ref{sec:cmodel}.

\section{Dynamical Model}\label{sec:Sys}

{In this section we  consider a set of $\mathcal{N}$ interacting agents, labeled $i=1,...,\mathcal{N}$. For each agent, we consider two types of (interdependent) dynamics: the robot spatial dynamics and the oscillator dynamics. The spatial dynamics of each robot can be obtained by its equation of motion, while the oscillators are characterized by either periodic or chaotic dynamics. The full state ${\bf{s}}_i(t)$ of agent $i$ can then be written as ${\bf{s}}_i(t)=[{\bf{p}}_i(t)^T,{\bf{v}}_i(t)^T,{\bf{x}}_i(t)^T]^T$, where ${\bf{p}}_i(t)$ and ${\bf{v}}_i(t)$ are the spatial position and velocity of robot $i$, respectively,  and ${\bf{x}}_i(t)$ is the state of oscillator $i$. The advantage of considering the oscillator dynamics is that, as we will see, these oscillators can be used as sensors on board of each robot, \eg they can provide information on the status of the local connectivity of each agent.}

\begin{figure}[h!]
\centering
\includegraphics[width=0.6\textwidth]{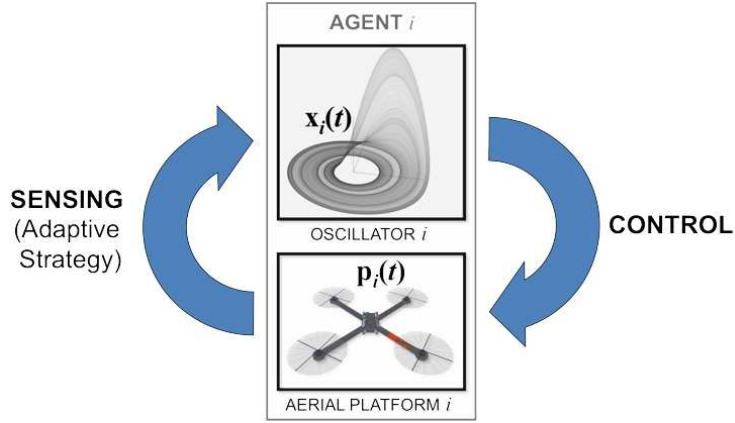}\label{agentf}
\caption{Each individual agent is composed of a robotic platform connected to a chaotic oscillator. The chaotic oscillator acts as a \emph{sensor} as it provides information on the distance to other platforms and on the presence of obstacles. This information can be fed back to \emph{control} the position of the platform. }
\end{figure}

 {Figure 2 shows an individual agent composed of a robotic platform connected to a chaotic oscillator. The chaotic oscillator acts as a \emph{sensor} as it provides information on the distance to the other platforms and the presence of obstacles. This information can be fed back to \emph{control} the position of the platform. While some preliminary results on the sensing strategy are contained in \cite{Bezzo+ACC:13}, to the best of our knowledge the control problem has not been investigated. A model for the dynamics of the robotic platforms and their communication and sensing capabilities is presented in Secs.\ \ref{sec:Robotmodel}, \ref{sec:Robotmodel2}, \ref{sec:cmodel}, \ref{Sec:VN}, and \ref{Sec:CN}. The synchronization dynamics of the oscillators is described in Sec.\ \ref{sec:Chaos}. In Sec.\ \ref{sec:AS} we present an example of application of an adaptive strategy to a network of coupled mobile platforms and in Sec.\ \ref{Sec:control} we present a specific example of \emph{decentralized formation control}.}

\subsection{Mobile Robot Dynamics}\label{sec:Robotmodel}

The dynamics of each robotic agent $i$ is approximated
by the following model,
\begin{eqnarray}\label{eq:doubleintegrator}
\dot {\mathbf p}_i &=& \mathbf{v}_i, \\
\dot {\mathbf v}_i &=& \mathbf{u}_i, \quad i= [1,\ldots,\mathcal N]  \nonumber,
\end{eqnarray}
where ${\mathbf p}_i = [y_i, z_i]^T \in \Re^2$ is the position
vector of  agent $i$ in the $(y, z)$ plane relative to a base station, ${\mathbf v}_i \in \Re^2$ and ${\mathbf
u}_i \in \Re^2$ denote the velocity and acceleration (control
input) vectors, respectively, for each agent $i \in \mathcal N$. The
 workspace $\mathcal{W}$ in which the agents operate is populated with $N_O$ fixed
polygonal obstacles $\{{O}_1,\ldots,{O}_{N_O}\}$,
whose geometries and positions are assumed unknown to the agents.

\subsection{Sensing}\label{sec:Robotmodel2}

In describing the mobile agents, we include some sensing capabilities inspired by real world devices to allow several functionalities.
In order to avoid the obstacles we model a ray field of view, similarly to a laser range finder footprint, and we create a repulsive potential that appears only when an obstacle is detected. We envision a scenario in which the agents have limited knowledge of the environment and the obstacles are detected locally when the robots move to certain locations of the workspace. Once an obstacle is detected, we define a local workspace repulsive potential field $W_{O,i}$ \cite{spong2006robot},  which approaches infinity as the robot approaches the obstacle, and goes to zero if the robot is at a distance greater than $\rho_{0}$ from the obstacle,
%
\begin{equation}
W_{O,i}=\left\{
    \begin{array}{ll}
        \frac{1}{2} \eta_{i} \left(\frac{1}{\rho({\mathbf p}_{i})}-\frac{1}{\rho_{0}}\right)^2& \textrm{if $\rho({\mathbf p}_{i}) \leq \rho_{0}$} \\
        0 & \textrm{if $\rho({\mathbf p}_{i}) > \rho_{0}$}
  \end{array}\right.,\label{eq:conob}
\end{equation}
where $\rho({\mathbf p}_{i})$ is the distance between agent $i$ and any detected obstacle in the workspace and $\eta_{i}$ is a positive constant.

Then, the repulsive force is equal to the negative gradient of $W_{O,i}$, \ie for $\rho({\mathbf p}_{i}) \leq \rho_{0}$ it is given by
\begin{equation}\label{eq:forceobst}
F_{O,i}=\eta_{i} \left(\frac{1}{\rho({\mathbf p}_{i})}-\frac{1}{\rho_{0}}\right)\frac{1}{\rho({\mathbf p}_{i})^2}\nabla_{{\mathbf p}_{i}}{\rho({\mathbf p}_{i})},
\end{equation}
where $\nabla_{{\mathbf p}_{i}}{\rho({\mathbf p}_{i})}=[\frac{\partial \rho}{\partial y_i}, \frac{\partial \rho}{\partial z_i}]^T$ is the gradient of the distance between agent $i$ and the detected obstacle.
Note that for ease of discussion, in this paper we consider simple scenarios with convex polygons. Therefore, we do not examine cases in which agents can get stuck in local minima. However, local minima can be avoided by using either random-walk or potential based techniques \cite{lee2010random}.


\subsection{Communication Model}\label{sec:cmodel}

Our work is motivated by the use of wireless radio and optical communication.
The signal propagated from a
transmitter (Tx) to a receiver (Rx) can be decomposed into three separated and
well-known characteristics: $path$ $loss$, $shadowing$, and
$multipath$ \cite{goldsmith2005wireless}. Path loss is caused by
dissipation of the radiated power from the transmitter and by
effects of the propagation channel. Shadowing is caused by obstacles between the transmitter and the receiver that create reflection, scattering, absorption and attenuation of the propagated signal. Finally, the constructive and destructive addition of multipath
components create rapid fluctuations of the received signal strength over short periods of time.
%

Path loss among all the above three modes is the only non random behavior because it depends on the distance separation between the transmitter and the receiver. Shadowing and multipath are random processes and require a finer mathematical description. Because of their nature, in this paper we model shadowing and multipath as noise.

In general a simplified path loss model can be built to capture the
essence of signal propagation. The following equation
\begin{equation}
P_r=P_t k \left[\frac{d_0}{d}\right]^{\tau}+\beta,
\label{eq:generalpathloss}
\end{equation}
represents a generalized approximation of a real channel, where $P_t$ is the transmitted power and $P_r\left(d\right)$ is the
received power which is function of the separation $d$ between Tx and Rx, $k$
is a constant that depends on the antenna characteristics and
channel attenuation, $d_0$ is a reference distance, $\tau$ is the path loss exponent, and $\beta$ is a {parameter} that takes into consideration the random effects due to shadowing and multipath.
Equation \eqref{eq:generalpathloss} can be simplified as
\begin{equation}
P_r=\frac{K}{d^{\tau}}+\beta,
\label{eq:generalpathloss}
\end{equation}
where $K=P_t k {d}^{\tau}_0$, assuming $P_t$ constant. Thus the received signal strength falls off in inverse proportion to the $\tau^{\text{th}}$ power of the distance $d$ between the transmitter and the receiver. The parameter $\tau$ is called the path loss exponent, which depends on
the environment  \cite{goldsmith2005wireless}. {For simplicity in this work we assume that $\beta$ is negligible and consider only path-loss effects as the predominant dissipative source,}
\begin{equation}
P_r\propto \frac{1}{d^{\tau}}.
\label{eq:generalpathlossprop}
\end{equation}

Finally, our weighted adjacency matrix will evolve in time according to this information on the received signal. In what follows we assume that all the nodes act as both transmitters and receivers and based on Eq.\ (\ref{eq:generalpathlossprop}), we take 
\begin{equation} \label{aijt}
A_{ij}(t)=A_{ji}(t)=  \left\{ \begin{array} {ccc} {P_{ij}(t)=\norm{\mathbf{p}_i(t) - \mathbf{p}_j(t)}_2^{-\tau},} \quad  & \mbox{if} \quad {(v_i,v_j) \in \mathcal{E}}, \\ 0,  \quad & \mbox{otherwise.} \end{array} \right.
\end{equation}
The particular case of $\tau=0$ corresponds to an unweighted graph, for which all the existing edges have associated weight equal $1$. For the remainder of this paper, without loss of generality we will set $\tau=2$.

\subsection{Virtual Leader Potential} \label{Sec:VN}

In this work we assume that the multi-robotic network is guided by a virtual leader agent, often referred in the literature as a {\em pin} \cite{pin} that acts as a mobile attractive potential.
Let $\zeta : \Re^n \rightarrow \Re$ be the leader potential function. With a proper design of $\zeta(\mathbf p)$, the mobile agents can be attracted or repulsed to certain areas of the environment. The case $\zeta(\mathbf p)$ constant represents an isotropic scenario or, in other words, an equipotential area in the environment \cite{gazi2004stability}.

We define a time varying attractive potential centered in the virtual leader position.
For the sake of simplicity, we use the following exponential weight function to represent the leader attractive function
\begin{equation}\label{eq:gaussian}
\zeta(\mathbf{p}_{i})=B_{\zeta}-\frac{A_{\zeta}}{2}\,e^{-\frac{\| \mathbf{p}_{i}-c_{\zeta}(t)\|^2}{\ell_{\zeta}}},
\end{equation}
where $A_{\zeta}$ and $\ell_{\zeta}$ are positive constants, $c_{\zeta}(t)\in \Re^{2}$ and $B_{\zeta}\geq A_{\zeta}/2 $. The term $\ell_{\zeta}$ controls the shape of the exponential function and $c_{\zeta}(t)$ is the center of the time varying user attractive function.
The gradient of \eqref{eq:gaussian} $\nabla_{\mathbf{p}_{i}}\zeta(\mathbf{p}_{i})=[\frac{\partial \zeta}{\partial y_i}, \frac{\partial \zeta}{\partial z_i}]^T$ is
\begin{equation}\label{eq:gradgauss2}
\nabla_{\mathbf{p}_{i}}\zeta(\mathbf{p}_{i})=\frac{A_{\zeta}}{\ell_{\zeta}}(\mathbf{p}_{i}-c_{\zeta}(t))\,e^{-\frac{\| \mathbf{p}_{i}-c_{\zeta}(t)\|^2}{\ell_{\zeta}}}.
\end{equation}

\subsection{Controller} \label{Sec:CN}

By assembling together the pieces described in the previous sections, we obtain the overall control law for the agents,
\begin{align} \label{eq:controller}
\ddot {\mathbf p}_{i} &=F_{O,i}-\nabla_{{\mathbf p}_{i}}\zeta({\mathbf p}_{i}).
 \end{align}

We assume that at the initial time, the network is fully connected and thus the only zero-entries for the weighted adjacency matrix are on the main diagonal.
If 
 the line-of-sight pathway between two agents is interrupted by an obstacle, that connection is assumed lost and thus the weighted adjacency matrix  has off-diagonal zero entries, too.

\subsection{Synchronization Dynamics of Chaotic Oscillators}\label{sec:Chaos}

We assume that each agent $i=1,...,\mathcal{N}$ is equipped with a nonlinear oscillator, whose dynamics is described by
\begin{equation}\label{net}
\dot {\mathbf x}_i(t) = F({\mathbf x}_i(t)) + \gamma {\mathbf \Gamma} [\sigma_i(t) r_i(t)-H({\mathbf x}_i(t))],
\end{equation}
 where, ${\mathbf x}_i(t)$  is the $q$-dimensional state of oscillator $i=1,...,\mathcal{N}$;  $F(x)$ determines the dynamics of an uncoupled ($\gamma \rightarrow 0$) system (hereafter assumed chaotic), $F:R^q \rightarrow R^q$; $H(x)$ is a scalar output function, $H:R^q \rightarrow R$. We take ${\mathbf \Gamma}$ to be a constant $q$-vector, ${\mathbf \Gamma}=[\Gamma_1,\Gamma_2,...,\Gamma_q]^T$ with $\sum_i \Gamma_i^2=1$ and the scalar $\gamma$ is a constant characterizing the strength of the coupling.
 The  scalar signal each node $i=1,...,\mathcal{N}$ receives from the other nodes in the network is
\begin{equation}\label{ri}
r_i(t)=\sum_j A_{ij}(t) H({\mathbf x}_j(t)). 
\end{equation}
{and $\sigma_i(t)$ is a parameter that we assume can be adaptively evolved at each node $i=1,...,\mathcal{N}$ (according to the adaptive strategy presented in Sec.\ \ref{sec:AS}). We proceed under the assumption that the transmission and measurement times for the received signal $r_i(t)$ are negligible. The effects of transmission delays along different paths $i \leftrightarrow j$ on the adaptive strategy have been studied in \cite{SOTT2}.}

We wish to emphasize that the entries of the adjacency matrix $A_{ij}(t)$ evolve in time based on both the relative positions of the mobile robots and the presence of obstacles along the line-of-sight pathways of the signals exchanged by the oscillators. We note that if the following condition is satisfied
 \begin{equation}
 \sigma_i(t)=\left[\sum_j A_{ij}\right]^{-1}, \label{goal}
 \end{equation}
 then Eq. (\ref{net}) admits a synchronized solution,
 \begin{equation}
 {\mathbf x}_1(t)={\mathbf x}_2(t)=...={\mathbf x}_{\mathcal N}(t)={\mathbf x}_s(t), \label{eq:xs}
 \end{equation}
  where ${\mathbf x}_s(t)$ satisfies
  \begin{equation}
  \dot{{\mathbf x}}_s(t)=F({\mathbf x}_s(t)), \label{five}
  \end{equation}
 which corresponds to the dynamics of an uncoupled system.

We note that Eqs. (\ref{goal}), (\ref{eq:xs}), and (\ref{five}) provide the desired (synchronized) solution for the system of equations (\ref{net}) and (\ref{ri}). However, other attractors for the dynamics may exist. This is discussed in \cite{SOTT}, where the importance of properly choosing the initial conditions of the oscillators variables ${\mathbf x}_i(0)$ and of the internal variables $m_i(0)$ and $n_i(0)$ is discussed (see in particular Fig.\ 2 and the accompanying discussion in the text of Ref.\ \cite{SOTT}).

Stability of the synchronous solution \eqref{eq:xs} for the case that the network topology is fixed and known has been studied in \cite{Pe:Ca}. In this paper we are interested in the case that the network is time varying and that the strength of the signal received by each oscillator is unknown and has to be adaptively computed.
In the following section we present an adaptive technique based on synchronization of chaotic systems, which we will successfully employ to track changes of the network connectivity.

\section{Adaptive Strategy}\label{sec:AS}

In what follows we regard $\sigma_i(t)$ as an internal adaptive parameter at each node $i=1,...,\mathcal{N}$. We propose a decentralized adaptive strategy to independently evolve $\sigma_i(t)$ at each node in order  to achieve dynamical approximate satisfaction of condition (\ref{goal}).
We regard the $A_{ij}(t)$ as unknown at each node $i$, while the only external information available at node $i$ is its received signal (\ref{ri}). The goal of the adaptive strategy is to adjust $\sigma_i(t)$ to maintain synchronism in the presence of slow and unknown time variations of the quantities $A_{ij}(t)$. A similar adaptive strategy was proposed first in \cite{SOTT,SAS} and experimentally tested for a static network of coupled opto-electronic oscillators in \cite{EXP,EXP2}.

We assume that each node implements the adaptive strategy in a decentralized way. Thus at each system node $i$, we define the exponentially weighted synchronization error $\psi_i=<(\sigma_i r_i -H({\mathbf x}_i))^2>_{\nu}$, where
 \begin{equation}
 <G(t)>_{\nu}=\int^t G(t') e^{-\nu (t'-t)} dt'
 \end{equation}
 and $\nu^{-1}$ is the time-window over which the average is performed. We then seek to evolve $\sigma_i(t)$ to minimize this error, based on the following gradient-descent relation,
 \begin{equation}\label{GD}
 \dot{\sigma}_i=-\kappa \frac{\partial \psi_i}{\partial \sigma_i},
 \end{equation}
where $\kappa$ is a positive scalar.  Following \cite{SOTT}, we assume that the dynamics of the adaptation process (\ref{GD}) is fast. Thus taking $\kappa \rightarrow \infty$,  we have that the right-hand-side of Eq.\ (\ref{GD}) rapidly converges to zero, corresponding to minimizing the potential. Hence, by setting ${\partial \psi_i}/{\partial \sigma_i}$ to zero, we can replace (\ref{GD}) by the following equations
\begin{equation}
\sigma_i(t)=\frac{<H({\mathbf x}_i(t)) r_i(t)>_{\nu}}{<r_i(t)^2>_{\nu}}=\frac{m_i(t)}{n_i(t)}. \label{sigma}
\end{equation}
 We now exploit the property that ${d<G(t)>_{\nu}}/{dt}=-\nu <G(t)>_{\nu}+G(t)$, from which we obtain that the numerator and the denominator on the right hand side of Eq. (\ref{sigma}) satisfy the differential equations
\begin{subequations} \label{quattro}
\begin{align}
\dot{m}_i(t)= & -\nu m_i(t)+ r_i(t) H({\mathbf x}_i(t)), \label{quattroa}\\
\dot{n}_i(t)=  & -\nu n_i(t)+ r_i(t)^2. \label{quattrob}
\end{align}
\end{subequations}

Since the dynamics of the couplings $A_{ij}$'s is imagined to occur on a timescale which is slow compared to the other dynamics in the network, we can approximate  $A_{ij}(t)$ as constant $A_{ij}$. 
In any practical situation, fulfillment of this assumption can be obtained by choosing the dynamics of the individual oscillators to be fast enough, \ie so as to verify the condition
\begin{equation}\label{TS}
T_x < {\nu}^{-1} \ll T_A,
\end{equation}
where $T_x$ and $T_A$ are the timescales on which the individual oscillators and the network  connections evolve, respectively.

 By assuming satisfaction of (\ref{TS}), we note that  Eqs. (\ref{net}), (\ref{sigma}), and (\ref{quattro}) admit a {\em synchronized solution}, given by Eqs. \eqref{eq:xs}, (\ref{five}), and
\begin{subequations}\label{cinque}
\begin{align}
\dot{m}_i^s= & -\nu m_i^s +(\sum_j A_{ij}) H({\mathbf x}^s)^2, \quad i=1,...,\mathcal N,  \label{cinqueb}\\
\dot{n}_i^s= & -\nu n_i^s+(\sum_j A_{ij})^2 H({\mathbf x}^s)^2, \quad i=1,...,\mathcal N. \label{cinquec}
\end{align}
\end{subequations}
If the synchronization scheme is stable, we expect that the synchronized solution \eqref{eq:xs}, (\ref{five}), and (\ref{cinque}) will be maintained under slow time evolution of the couplings $A_{ij}(t)$.

Stability of the adaptive synchronization strategy, given by Eqs.\ (\ref{net}), (\ref{sigma}), and (\ref{quattro}) about the synchronized solution \eqref{eq:xs}, (\ref{five}), and (\ref{cinque}), has been investigated in \cite{SAS}. In particular, it was found that if (\ref{TS}) holds, then stability of the synchronized solution depends on the pairs $(\nu, \xi_i)$, $i=1,...,(\mathcal{N}-1)$ where $\xi_i=\gamma (1-\alpha_i)$, and $\{\alpha_i\}_{i=1}^{\mathcal N}$ are the eigenvalues of the normalized network adjacency matrix $A'=\{A'_{ij}\}$, $A'_{ij}=(\sum_j A_{ij})^{-1} A_{ij}$, excluding the one eigenvalue $\alpha_{\mathcal N}=1$. Note that if the matrix $A$ is symmetric (which is always the case in our application as we defined $A_{ij}=A_{ji}$ to be a function the distance between node $i$ and $j$) the spectrum of the matrix $A'$ is real. In order to show this property, we first rewrite the matrix $A'$ as $A'=K^{-1} A$,
where $K$ is a diagonal matrix with $K_{ii}= \sum_j A_{ij}$. Then we note that by left-multiplying by $K^{1/2}$ the eigenvalue equation $A' {\textbf c} = \lambda {\textbf c}$ we obtain $(K^{-1/2} A K^{-1/2}) {\textbf c'} = \lambda {\textbf c'} $, where ${\textbf c'} = K^{1/2} {\textbf c} $. Because $K^{-1/2} A K^{-1/2}$ is symmetric, if $A$ is, it follows that the eigenvalues of $A'$ are real.

In Ref.\ \cite{SAS} stability was numerically investigated for the case that the individual systems are R\"ossler chaotic oscillators \cite{Rossler}, $q=3$, $\mathbf{x}(t)=(x^{(1)}(t),x^{(2)}(t),x^{(3)}(t))^T$,
\begin{equation}
F({{\mathbf x}})=\left[\begin{array}{c}
    -x^{(2)}-x^{(3)}  \\
    x^{(1)} + a x^{(2)} \\
    b+ (x^{(1)}-c )  x^{(3)}
  \end{array} \right], \label{RC}
\end{equation}
with parameters $a=b=0.2$, and $c=7$, and with $H(x(t))=x^{(1)}(t)$, $\Gamma=[1,0,0]^T$, and $\gamma=1$.  For example, for $\nu=0.5$,
 we see from Fig.\ 3 of Ref.\ \cite{SAS} that the condition for stability is that $0.7\leq\xi_i\leq 4.8$, $i=1,...,(\mathcal{N}-1)$ (in Fig.\ 3 of Ref.\ \cite{SAS} the stability area is the area of the $(\nu,\xi)$-plane delimited by the thick solid level-curves).

\section{Simulation Results}\label{sec:Sim}

In this section we present some simulation results to demonstrate the applicability of the adaptive strategy described in Sec.\ 4. We present two numerical experiments.
In the first one, we consider a network composed of five agents that move in an obstacle populated environment and get disconnected and reconnected as the simulation evolves.
In the second one, we consider three agents as a special case for which it is possible to individually estimate all of the entries of the adjacency matrix. As we will show, for this case, we are able to dynamically reconstruct the time evolution of all the individual couplings between the agents in the system.

\subsection{Tracking of Variations in the Network Connectivity}

\begin{figure}[ht!]
\vspace{4pt}
\begin{center}
\subfigure[] {\includegraphics[width=0.238\textwidth]{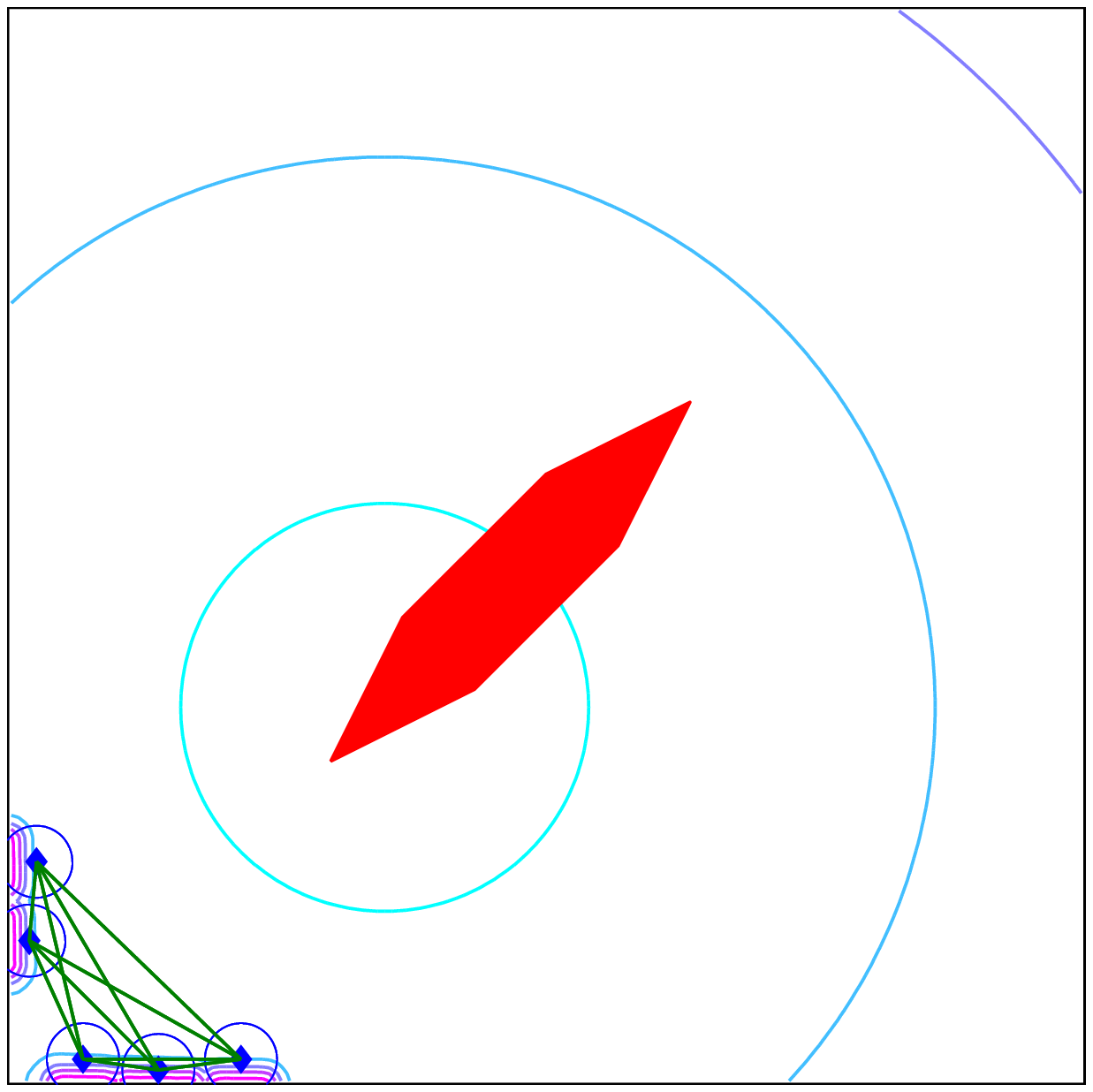}}
\subfigure[] {\includegraphics[width=0.238\textwidth]{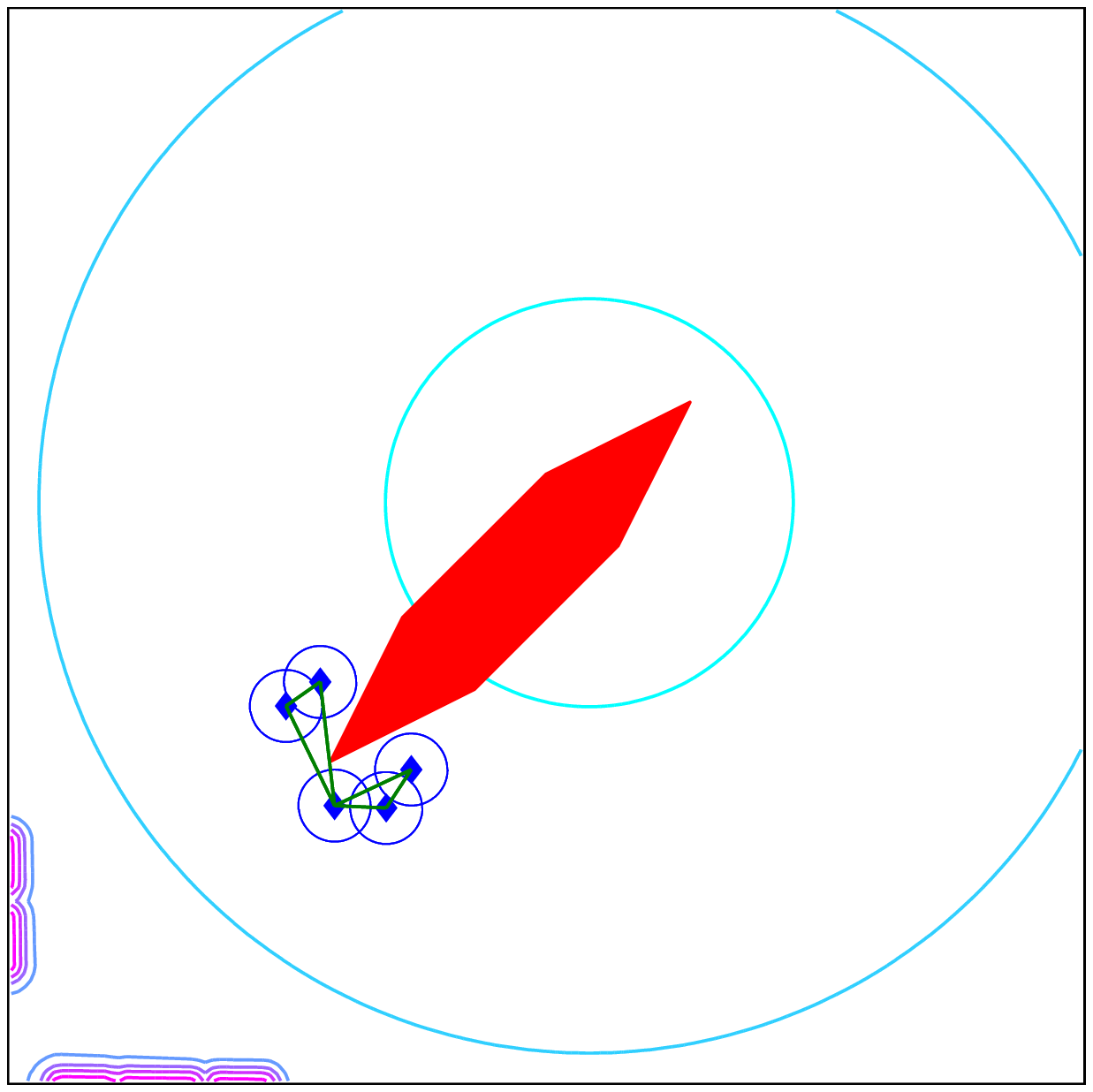}}
\subfigure[] {\includegraphics[width=0.238\textwidth]{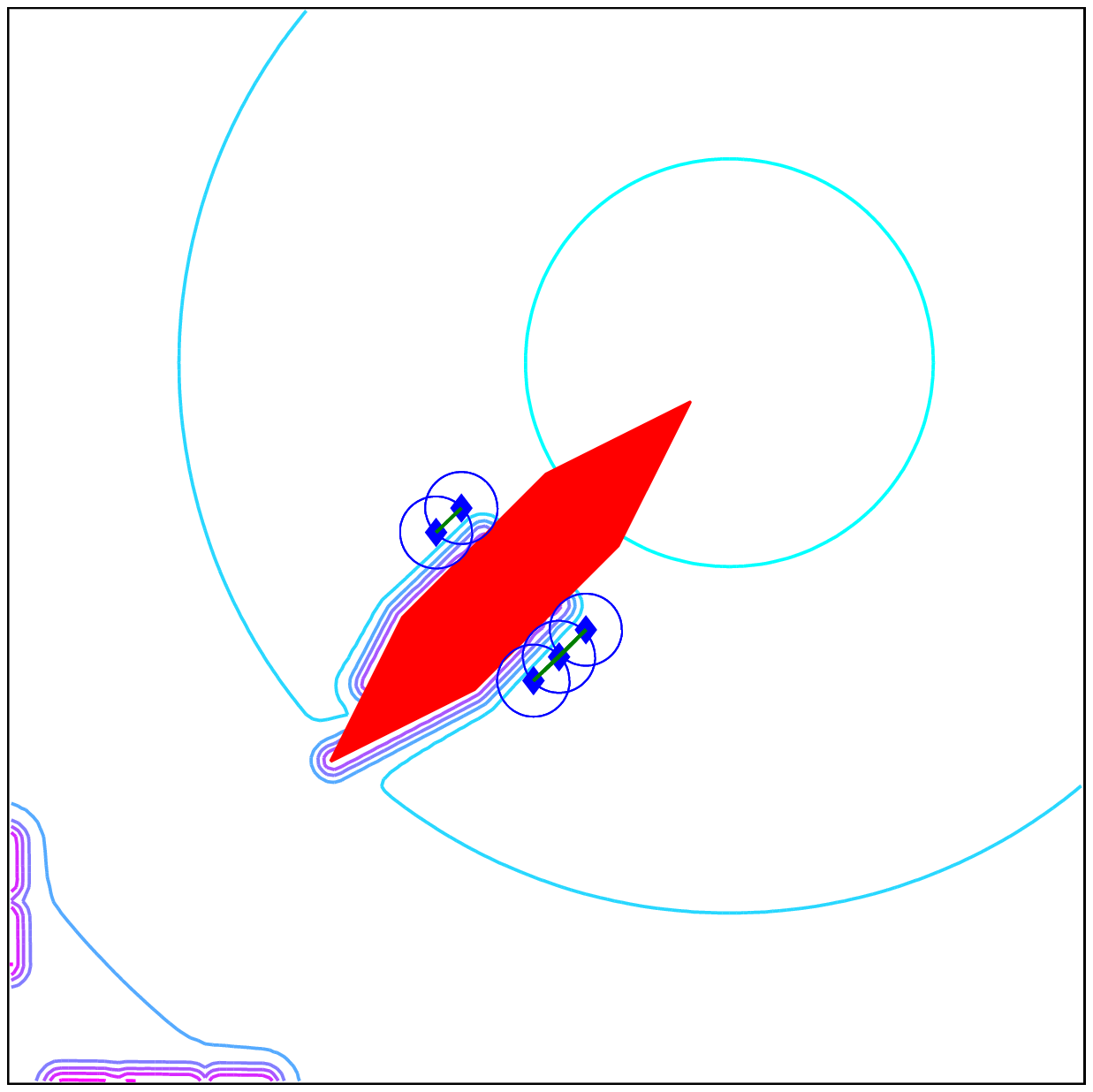}}
\subfigure[] {\includegraphics[width=0.238\textwidth]{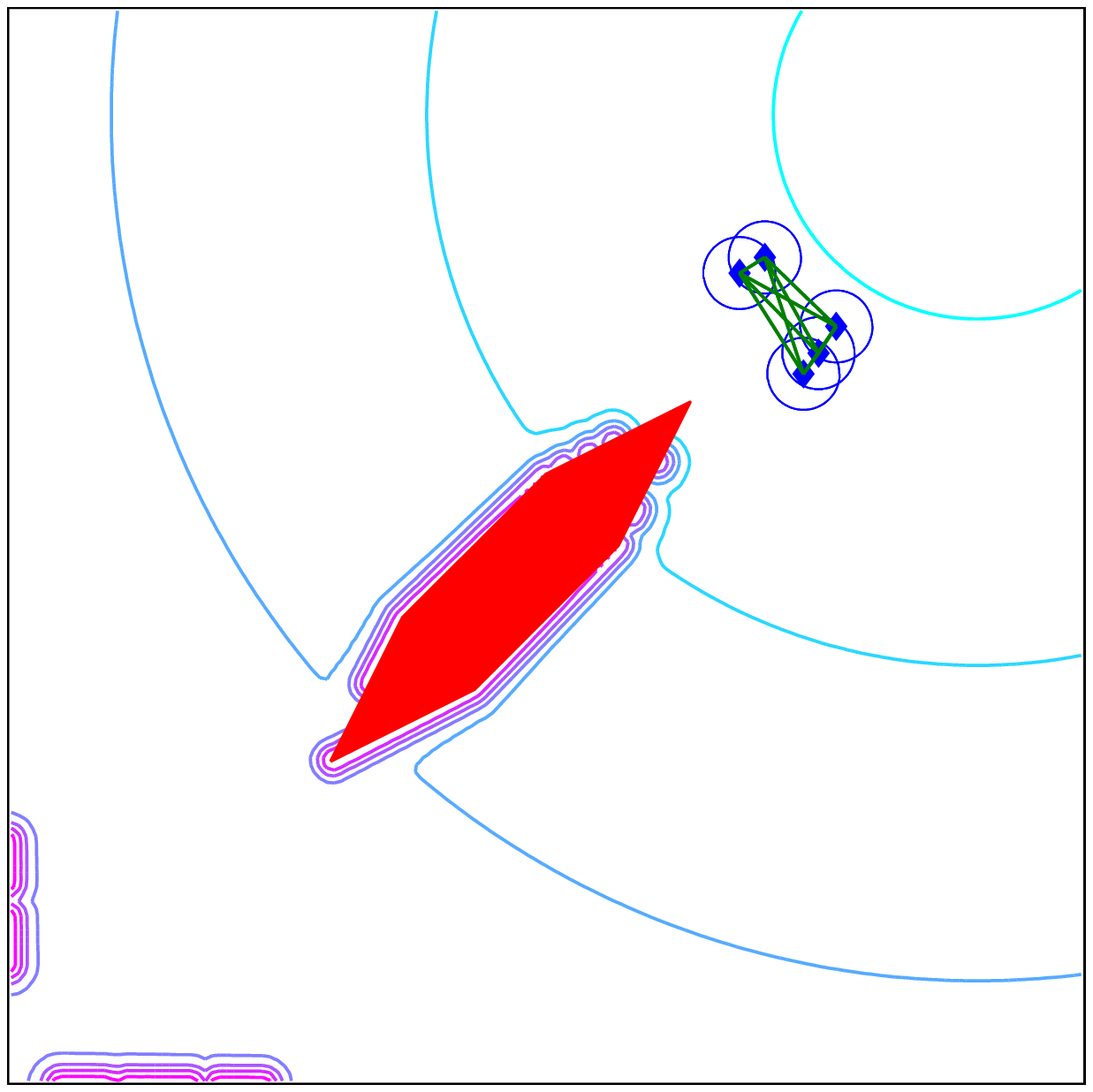}}
\vspace{-15pt}
\end{center}
\caption{Simulation  of five mobile robots moving in a cluttered environment. The lines drawn between agents represent line-of-sight communication links. The movie related to this simulation is available in the supplementary information (movie $1$).}
\label{fig: sim1}
\end{figure}

Our first numerical experiment is illustrated in Fig.~\ref{fig: sim1}.
We consider a network of five robots moving in a workspace with an obstacle in its center. The agents are guided by a virtual potential, described in Sec. \ref{Sec:VN}, towards a location in the environment. The virtual node is driving from the lower corner of Fig.~\ref{fig: sim1}(a) to the upper corner of the workspace (Fig.~\ref{fig: sim1}(d)). Because of the presence of the obstacle, in Figs.~\ref{fig: sim1}(b)(c) the network gets disconnected and is divided into two groups that reconnect  towards the end of the mission (Fig.~\ref{fig: sim1}(d)). The obstacle is detected locally by the agents as they move attracted by the potential of the virtual node.

\begin{figure}[ht!]
\centering
\includegraphics[width=0.485 \textwidth]{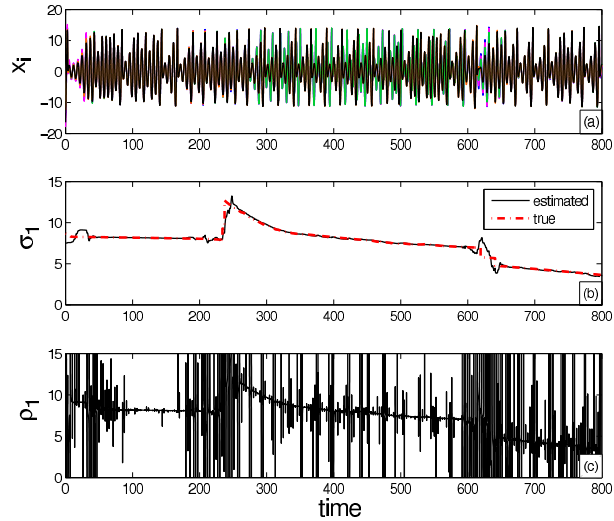}
\vspace{-0pt}
\caption{\label{fig:res} Dynamics of the oscillators for the simulation with five robots in Fig.~\ref{fig: sim1}. a) Plot of the state time evolution of the five oscillators. b) Comparison between the estimated and the true values of $\sigma$ for one of the robots. c) Plot of the ratio between the transmitted and the received signals for the same robot.}
\end{figure}

\begin{figure}[ht!]
\vspace{4pt}
\centering
\includegraphics[width=0.485 \textwidth]{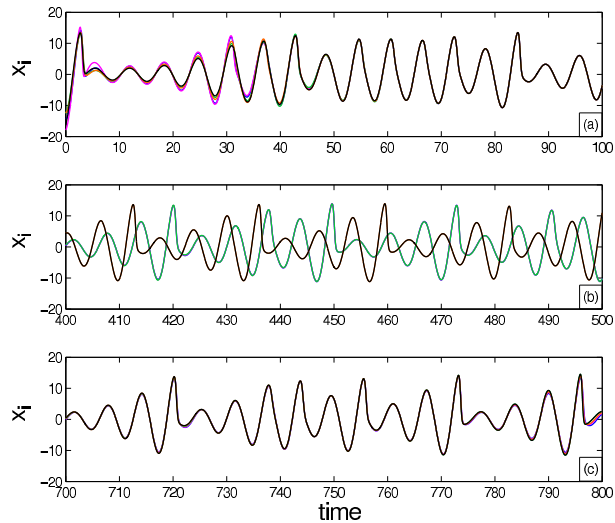}
\vspace{-0pt}
\caption{\label{fig:res2} Synchronization of chaotic oscillators during the experiment with five mobile agents in Fig.\ref{fig: sim1}. a) The five chaotic signals start not synchronized and synchronize around time 40. b) Two synchronized orbits created while the system is divided in two groups. c) Resynchronization of the entire group towards the end of the simulation.}
\end{figure}

The corresponding dynamics of the oscillators can be observed in Fig.~\ref{fig:res}. Specifically, Fig.~\ref{fig:res}(a) shows the time evolution of the five chaotic oscillators. Fig.~\ref{fig:res2} presents the results of Fig.~\ref{fig:res}(a) in finer detail and will be subject of discussion later on in this section. In the Fig.~\ref{fig:res}(b), the estimated value of $\sigma$ (solid dark line) for one of the robots is plotted in comparison with its true value (dashed red line). As can be noticed, after an initial transient, the agent is able to synchronize with the rest of the group (Fig.~\ref{fig:res2}(a)) and stays on a synchronous orbit until an event occurs. When a disconnection happens, the superimposed signal received by some of the agents decreases creating a jump in the value of $\sigma$ and a consequent loss of synchrony. For example, between time 200 and 300 in Fig.~\ref{fig:res} the network of 5 robots breaks down into two groups, as depicted in Figs.~\ref{fig: sim1}(b) and (c).

As the network disconnects into two groups, we observe from Fig.~\ref{fig:res2}(b) that each one of the two groups synchronizes on a distinct time evolution (the two times evolutions are visible in the central part of Fig.~\ref{fig:res}(a) but are better distinguishable in Fig.~\ref{fig:res2}(b)). This is due to the property of the individual oscillators of being chaotic, as slightly different initial conditions or small dynamical perturbations result in exponential divergence of the trajectories (a property known as sensitive-dependence on the initial conditions \cite{OB}).
Subsequently, the two groups reconnect between time  600 and 700 (see Fig.~\ref{fig: sim1}(d)). After that, the whole network reconnects and resynchronizes, as can be seen from the close-up in Fig.~\ref{fig:res2}(c).
By looking at $\sigma_1$ in Fig.~\ref{fig:res}(b) we are able to identify variations on the net received signal, from which information can be gathered on the local connectivity of the network and its time variation.
Lost links can be detected by an increase of this quantity, while addition of links through a drop of this quantity.

Each agent $i$ can also independently compute the quantity
\begin{equation}
\rho_i=\frac{H({\mathbf x_i})}{r_i},
\end{equation}
\ie the ratio between the transmitted and the received signals for agent $i$. This quantity is particularly significant as it can be independently computed by each individual node and provides useful information about the local connectivity of the network.
Fig.~\ref{fig:res}(c) shows the time evolution of $\rho_i$ for  agent $i=1$. This graph suggests that each agent can track changes in network topology by comparing the received signal with the transmitted one. As can be seen from Fig.~\ref{fig:res}(c), the quantity $\rho_1$ replicates the time evolution of $\sigma_1$, though rapid variations from the profile of $\sigma_1$ are present during the whole time-span of the simulation. These deviations are the result of perturbations away from the synchronization manifold, due to the presence of noise and to the time variations of the couplings $A_{ij}(t)$. 
Monitoring the frequency of these deviations can be useful to detect events such as deletion or aggregation of signals. In particular, when these events occur, the frequency of the deviations increases.


\vspace{-4pt}
\subsection{Connectivity Estimation}

\begin{figure}[h!]
\begin{center}
\subfigure[] {\includegraphics[width=0.238\textwidth]{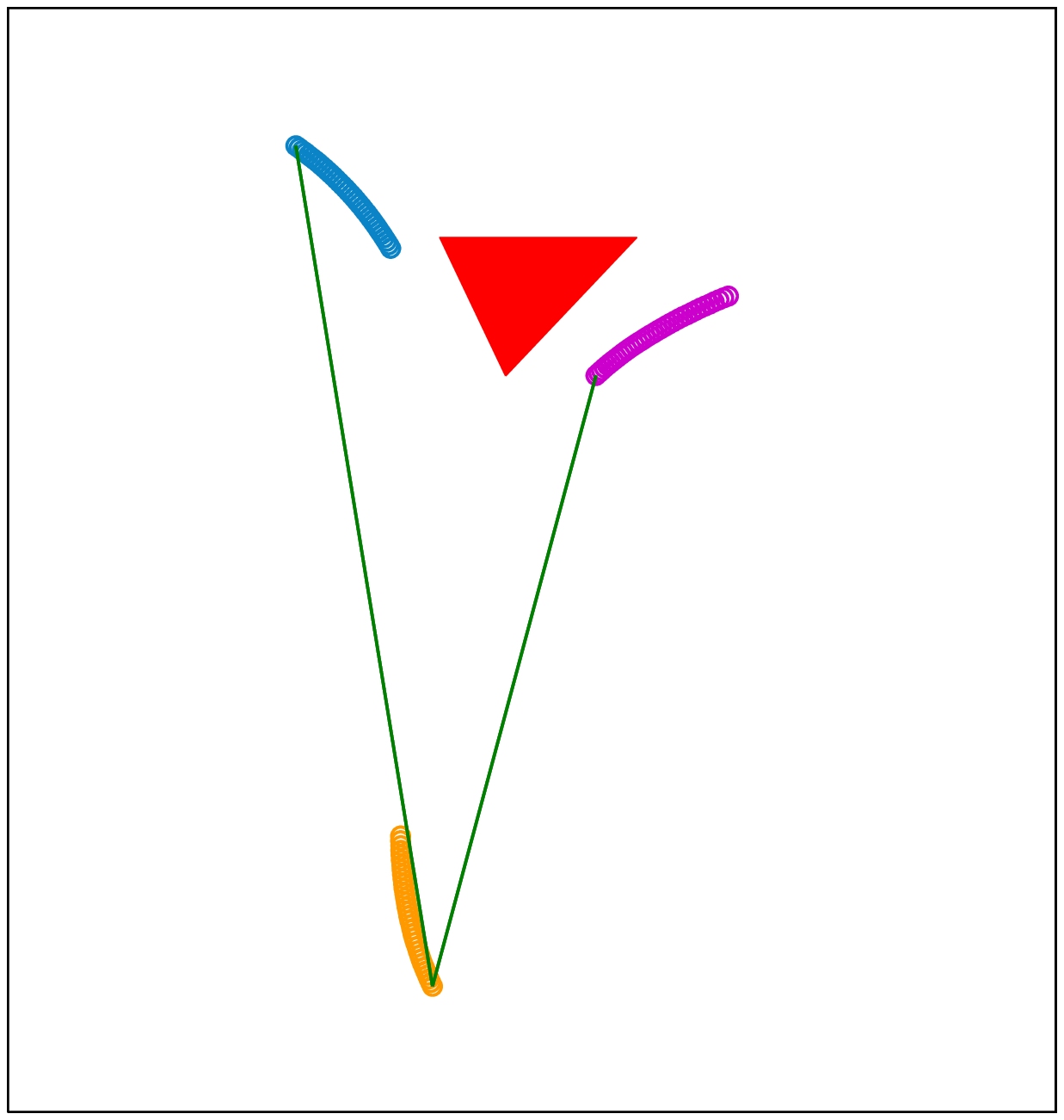}}
\subfigure[] {\includegraphics[width=0.238\textwidth]{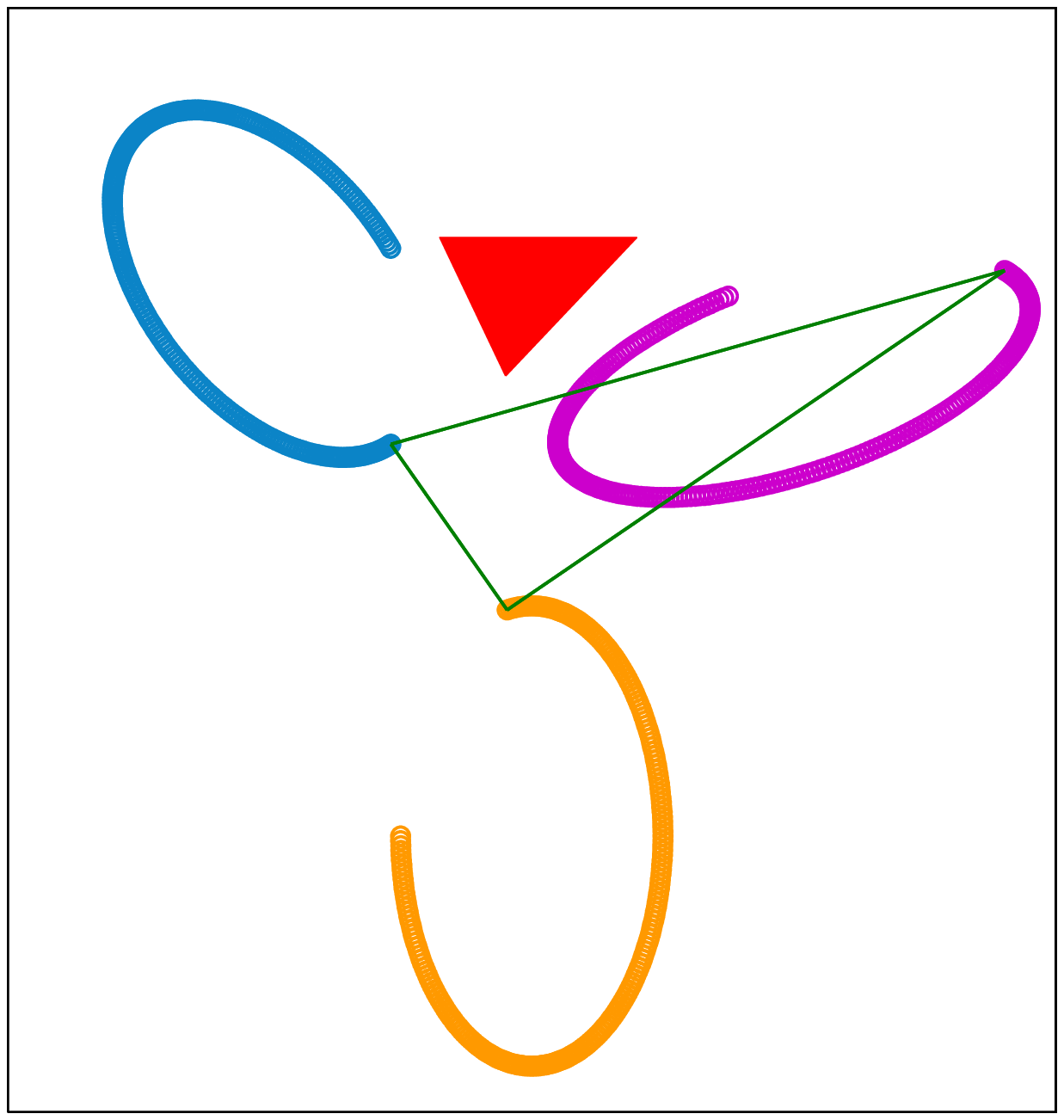}}
\end{center}
\vspace{-8pt}
\caption{Two snapshots of the simulation of three mobile robots moving along three elliptical curves. The triangular obstacle is responsible for discontinuities in the connectivity between the agents, \ie a connection is lost any time that a communication pathway crosses the obstacle. The movie related to this simulation is available in the supplementary information (movie $2$).}
\label{fig: sim2}
\end{figure}

As a second case study, we consider a network composed of ${\mathcal N}=3$ mobile robots moving in an obstacle populated workspace (see Fig.\ \ref{fig: sim2}). For this particular simulation, each platform is constrained to move along an elliptical trajectory at constant velocity. Moreover, each agent receives a signal that is a superimposition of maximum two chaotic signals, which makes the estimation of the individual links possible. 
In fact, for this case, from knowledge of $\sigma_1$, $\sigma_2$, and $\sigma_3$ (\ie assuming that each $i$ broadcasts information on $\sigma_i$ as well as on $\mathbf x_i$), it becomes possible for each agent to dynamically reconstruct the time evolution of the whole network adjacency matrix \cite{EXP2}.  The estimated values of the true couplings $A_{ij}$ can be obtained as $\hat{A}_{ij}=(\sigma_i^{-1}+\sigma_j^{-1}-\sigma_k^{-1})/2$, for $i,j = 1,\ldots, 3$ and $k\neq i\neq j$.

\begin{figure}[h!]
\centering
\includegraphics[width=0.485 \textwidth]{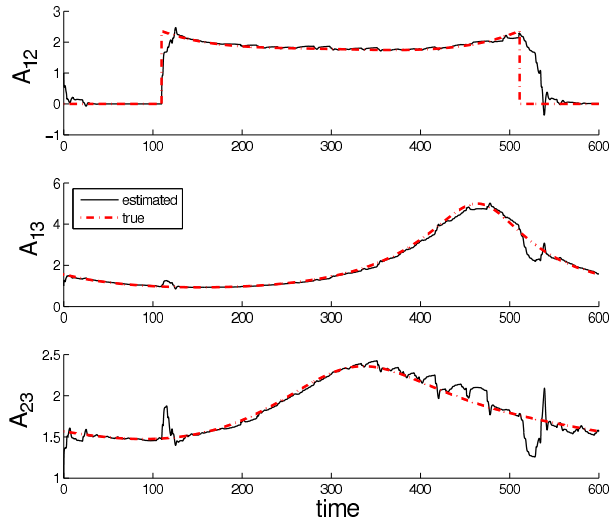}
\vspace{-0pt}
\caption{\label{fig:res3} Comparison between estimated values (solid lines) and true values (dashed lines) for the entries of the adjacency matrix obtained from the simulation with three robots in Fig.~\ref{fig: sim2}.}
\end{figure} 

Figure \ref{fig:res3} shows the comparison between the estimated and the true values of $A_{12}=A_{21}$, $A_{13}=A_{31}$, and $A_{23}=A_{32}$ (the true values are plotted as solid dark lines and their estimates as dashed red lines).
As can be seen, the proposed strategy performs well and allows each agent to both track changes of the network topology and  estimate the individual strengths of the signals received by the neighboring nodes. Each platform can then use this information, together with knowledge of the relation in Eq.\ (\ref{aijt}), to dynamically estimate the distances to the others and to detect the presence of obstacles along their communication pathways (see \eg Fig.~\ref{fig: sim2}.)

\section{Advantages of using chaotic oscillators} \label{Sec:Advantages}

In this section we discuss the importance of using chaotic oscillators in the estimation strategy. Ref.\ \cite{SOTT2} considers an adaptive strategy similar to the one described in this paper and points out the convenience of using chaotic oscillators with respect to other types of oscillators (\eg harmonic oscillators) in terms of the number of unknown quantities that each individual agent $i$ can independently extract from the received signal $r_i(t)$, $i=1,..,\mathcal{N}$. This also means that  the adaptive strategy could be used to estimate other parameters of interest besides the strengths of the connections, such as for example, some of the internal parameters of the chaotic system at a transmitter node (see \eg \cite{rr4,rr5}).
Hence, in terms of our application, by using chaotic systems,  we may be able to estimate internal parameters specific to certain types of robots (\eg to discriminate between aerial and ground agents, each encoded with a different parameter). Moreover,  chaotic signals appear as noise to any external party and thus provide improved security of the communication when conflicting  agents operate in the same environment,  such as in pursuit and evasion games.

In this paper, we have also shown another specific advantage of using chaotic oscillators, in terms of their property of being sensitively dependent on the initial conditions. In particular, as can be seen from Fig.\ 5(b), in the absence of coupling between the oscillators, small perturbations from the synchronous state grow unboundedly, whereas small perturbations would result in small deviations of the dynamics for periodic oscillators. This has immediate practical relevance for the case that the network gets disconnected and separates in two subgroups shown in Figs.\ 4 and 5. For example, by looking at the dynamical time-evolution of the nodes shown in Fig.\ 5(b),  it would be clear to an external observer that the network has disconnected in two separate subgroups, each of them following a distinct chaotic time-evolution. The same conclusion would be difficult to draw if periodic oscillators were used instead.

\section{Decentralized Formation Control} \label{Sec:control}

\begin{figure}[ht!]
\centering
\vspace{6pt}
\includegraphics[width=0.48 \textwidth]{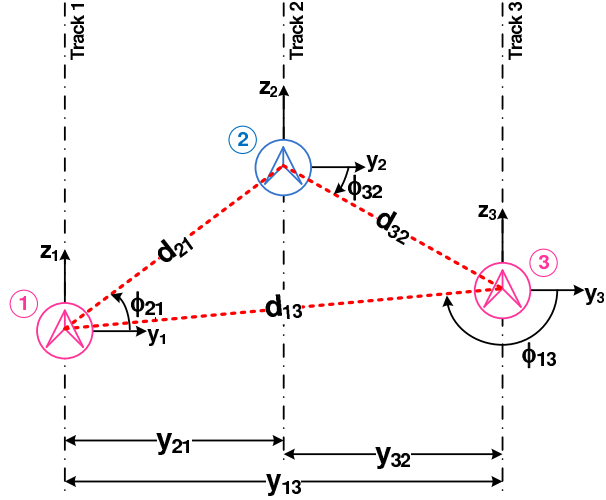}
\caption{\label{fig:sec8} The figure describes the problem considered in Sec.\ 7, namely a set of ${\mathcal N}=3$ platforms are constrained to move along imaginary tracks parallel to the $z$ direction, while trying to achieving and/or maintaining a desired formation. The formation is defined in terms of a set of desired relative distances $d_{ij}^d$ and desired bearing coordinates $\phi_{ij}^d$. We set the central platform labeled $2$ in the figure (in blue) to be the \emph{leader}, and the two outer platforms labeled $1$ and $3$ in the figure (in magenta) to be the \emph{followers}. This corresponds to setting $\ell=2$, \ie $\epsilon_1=\epsilon_3=1$, and $\epsilon_2=0$ in Eq.\ (24). }
\end{figure}

In Sec.\ \ref{sec:AS} we have proposed an adaptive strategy that has proven effective in producing a dynamical estimate of the distances between a set of coupled mobile platforms. In this section we go one step forward and propose to use the estimates obtained through implementation of the adaptive strategy  in order to maintain a certain formation between the mobile platforms. We derive and test a novel decentralized formation control algorithm,  based on the adaptive synchronization strategy described in Sec.\ \ref{sec:AS} for estimating the distances. The main difference with what we have done in Sec.\ \ref{sec:AS}, is that therein we assumed that the dynamics of the platforms affected the dynamics of the oscillators but not viceversa, while here the estimates produced by the adaptive strategy are fed back to the dynamics of the platforms, and in so doing, the loop is closed (see Fig.\ 2).

We discuss a specific example, in which the platforms move in an inhomogeneous and time-varying environment. An appropriate control strategy is then implemented in order to maintain the desired formation.
For this specific example, we consider that each platform is constrained to move along a line parallel to the $z$ axis, while its $y$ coordinate remains fixed. Say ${\mathbf p}_i = [y_i,z_i]^T \in \Re^2$ the position
vector of agent $i$ in the $(y,z)$-plane relative to a base station.  A graphical representation of our problem of interest is presented in Fig.\ \ref{fig:sec8}. The set of equations describing the dynamics is
 \begin{eqnarray}\label{eq:doubleintegrator}
\dot z_i(t) &=& {v^z}_i(t)+b_i(t),  \\
\dot { v^z}_i(t) &=& \epsilon_i {u}_i(t),  \nonumber
\end{eqnarray}
$i= 1,\ldots,\mathcal N$, where $v^z_i(t)$ is the velocity of particle $i$ along the $z$ direction, $b_i(t)$ is an external drag force acting on each platform $i$ at time $t$, representing the effect of the resistant fluid in which $i$ is moving, and $u_i(t)$ is a control input. The $\epsilon_i$ are binary quantities, \ie $\epsilon_i$ is equal $1$ ($0$) if platform $i$ is (is not) controlled. In what follows, we choose $\epsilon_i=(1-\delta_{i\ell})$, where $\delta_{ij}$ is the Kronecker delta, and the index $\ell$ indicates one platform which is chosen to act as a \emph{leader}. {In what follows we assume to have good control over the initial conditions of the oscillators variables ${\mathbf x}_i(0)$ and of the internal variables $m_i(0)$ and $n_i(0)$, $i=1,...,{\mathcal N}$.}

We set
\begin{equation}
b_i(t)=c_i+\frac{c_i}{2} \sin(\omega_i t +\theta_i),
\end{equation}
$i=1,...,{\mathcal N}$, where $c_i$ and $\omega_i$ are positive  numbers and $\theta_i$ represents an arbitrary phase.
The control input at node $i$ is chosen to be
\begin{equation}\label{ui}
u_i(t)=w \sum_{j\neq i} [\hat{d}_{ij} \sin(\phi_{ij})-d_{ij}^d \sin(\phi_{ij}^d)],
\end{equation}
$i=1,...,{\mathcal N}$, where $w$ is a positive gain, $\hat{d}_{ij}$ ($d_{ij}^d$) is the estimated (desired) Euclidean distance between platform $i$ and $j$, and $\phi_{ij}$ ($\phi_{ij}^d$) is the measured (desired) bearing coordinate between platform $i$ and $j$. Note that the quantities $\hat{d}_{ij}$ in Eq.\ \eqref{ui} are estimates of the unknown true distances $d_{ij}$.

In some applications, bearing measurements are easier to obtain than distance measurements. Indeed, this has motivated some work \cite{loizou2007biologically,bishop2010bearing} where formation control is obtained by using only bearing measurements. In what follows, we assume that precise direct measurements of the bearings are available, while the distances between the platforms are dynamically estimated. In particular, dynamical estimates of the unknown quantities $d_{ij}$ are obtained by using Eq.\ (\ref{aijt}), \ie by setting
\begin{equation}
\hat{d}_{ij}={\hat A}_{ij}^\tau,
\end{equation}
where ${\hat A}_{ij}$ is the estimate of the true coupling $A_{ij}$ provided by the adaptive strategy (see Sec.\ 5.2).

\begin{figure}[ht!]
\begin{center}
\subfigure[] {\includegraphics[width=0.40\textwidth]{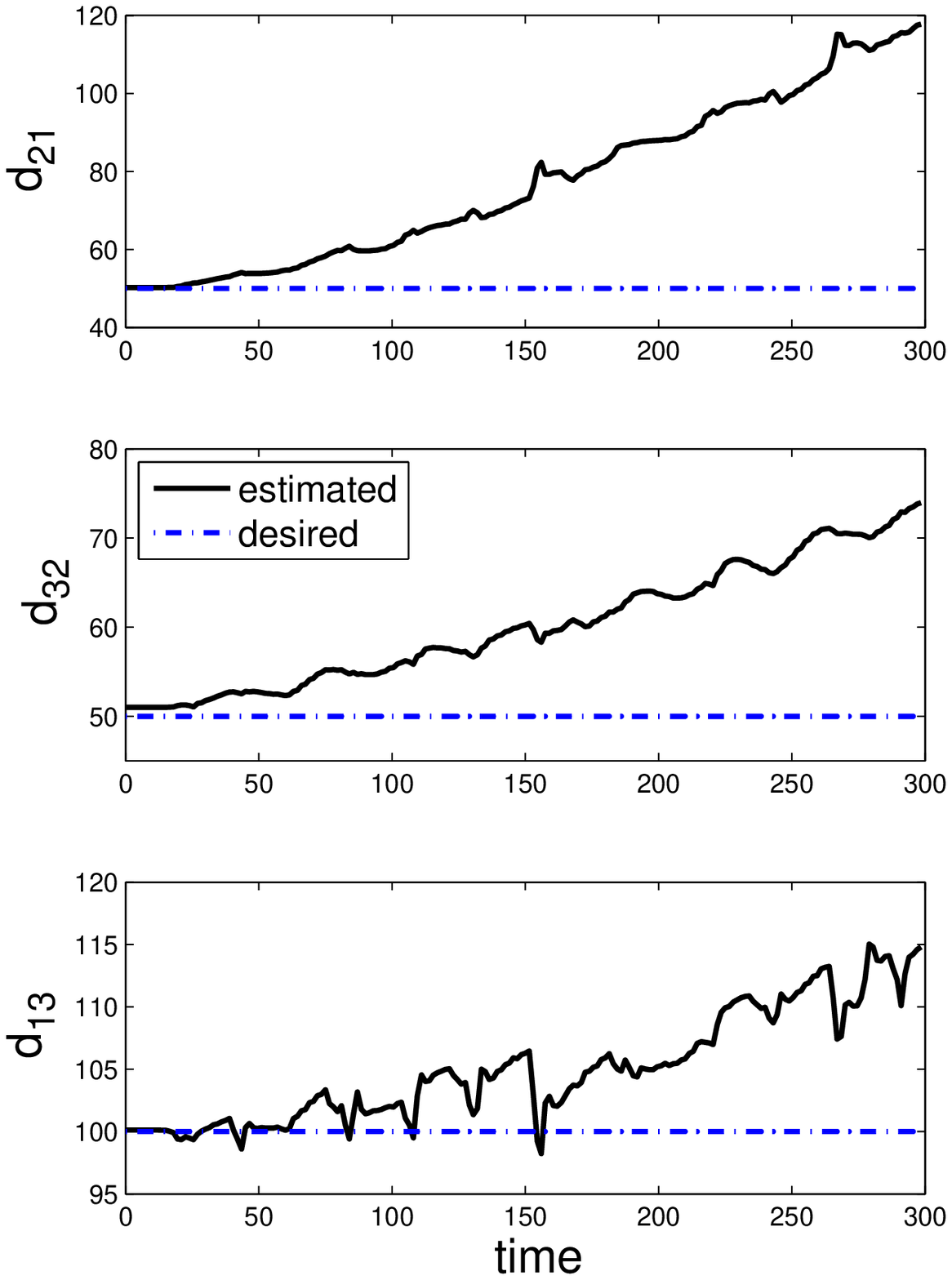}}
\subfigure[] {\includegraphics[width=0.40\textwidth]{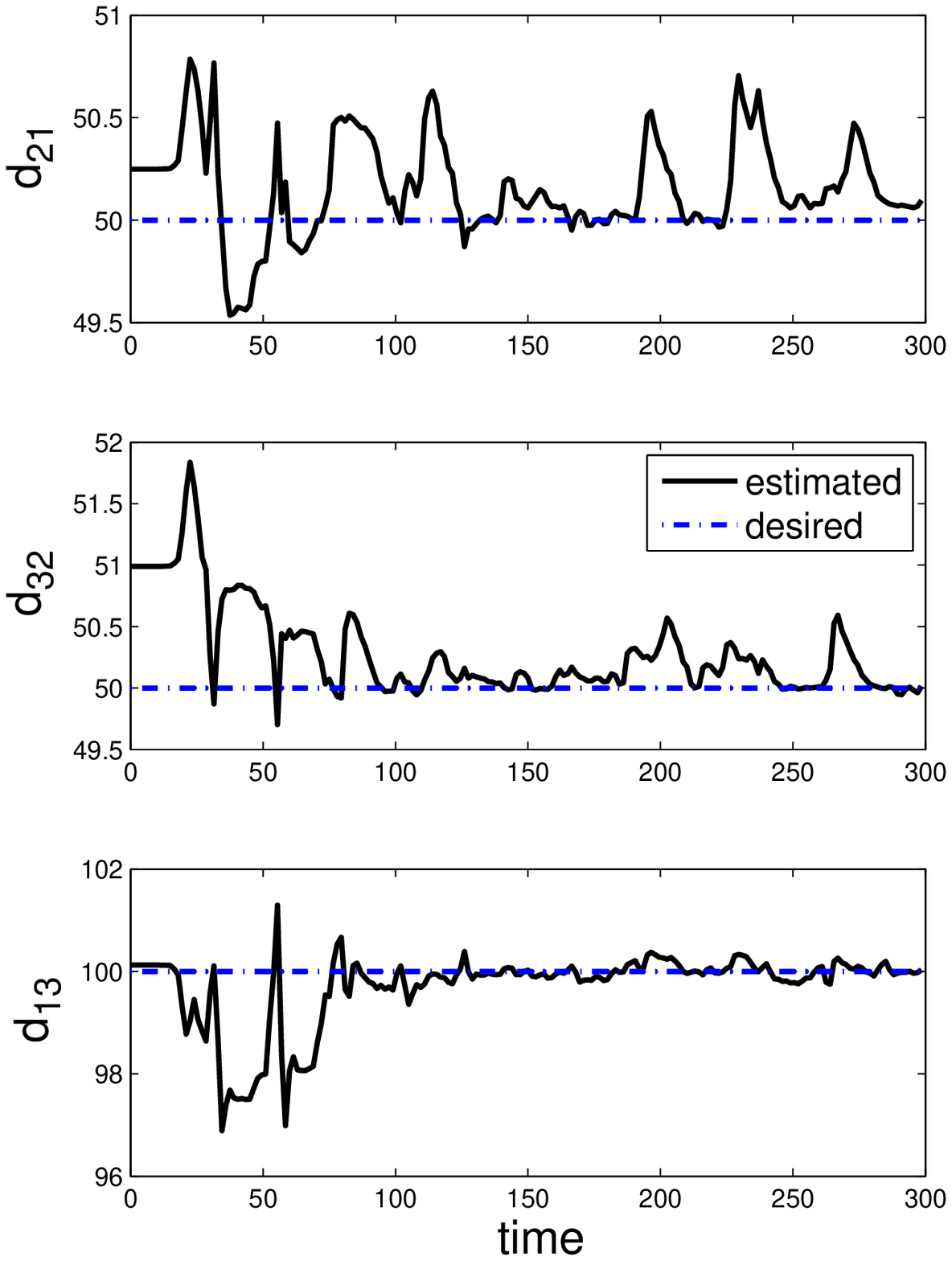}}
\end{center}
\vspace{-8pt}
\caption{The plots on the left-hand-side  represent a case where formation control is not implemented, \ie for which $w$ is set to $0$ in Eq. (\ref{ui}).   The movie related to this simulation is available in the supplementary information (movie $3$). The plots on the right-hand-side  represent a case where formation control is implemented, \ie for which $w$ is set to $0.1$ in Eq. (\ref{ui}).   The movie related to this simulation is available in the supplementary information (movie $4$). For the case on the right-hand-side, the desired formation is attained, with both the distances and the bearing coefficients approximately converging on their desired values. The parameters of the simulations are the following: $c_1=0.175$, $c_2=0.575$, $c_3=0.375$, $\omega_1=0.1256$, $\omega_2=0.1675$, $\omega_3=0.2094$, $\theta_1=\theta_2=\theta_3=0$, $d_{12}^d=50$, $d_{13}^d=100$, $d_{32}^d=50$, $\phi_{ij}^d=0$ for all the pairs $i,j$.}
\label{fig:FC}
\end{figure}

We run two numerical simulations, one for a case in which the formation control strategy  is implemented ($w=0.1$) and one for a case in which it is not ($w=0$). Fig.\ \ref{fig:FC} shows a comparison between the desired and estimated distances for the two cases, corresponding to $w=0$ (plots on the left-hand-side) and $w=0.1$ (plots on the right-hand-side), respectively.
The movies related to these simulations are available in the supplementary information (movies $3$ and $4$). Note that for $w>0$, the simulations are obtained by simultaneously implementing the estimation and control strategies. Additional plots showing the performance of the estimation strategy corresponding to the simulations shown in Fig. \ref{fig:FC} can be found online at {\em http://marhes.ece.unm.edu/index.php/PhysicaD}.

\begin{figure}[tb!]
\centering
\includegraphics[width=0.485 \textwidth]{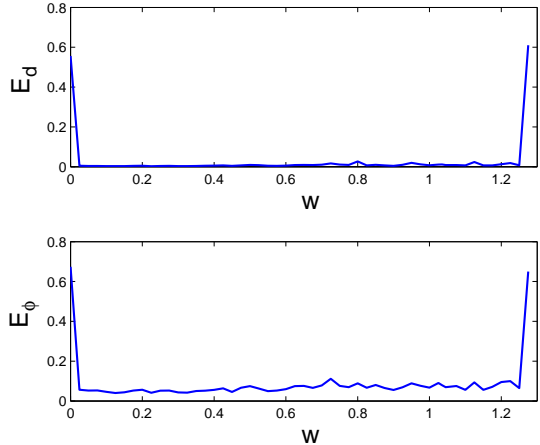}
\vspace{-15pt}
\caption{\label{fig:FC2} The distance error $E_d$ and the bearing error $E_\phi$ as a function of the control gain $w$. The parameters for this simulation are the same as in Fig.\ \ref{fig:FC}.  }
\end{figure} 

In order to characterize the performance of the above formation control strategy for different choices of the parameters, we introduce the following error measures, the distance error
\begin{equation}
E_d=100^{-1} (N(N-1)/2)^{-1} \int_{100}^{200} \sum_{i=1}^N \sum_{j=i+1}^N \frac{|d_{ij}(t)-d^d_{ij}(t)|}{d^d_{ij}(t)} dt,
\end{equation}
and the bearing error
\begin{equation}
E_\phi=100^{-1} (N(N-1)/2)^{-1} \int_{100}^{200} \sum_{i=1}^N \sum_{j=i+1}^N {|\phi_{ij}(t)-\phi^d_{ij}(t)|} dt.
\end{equation}

Figure \ref{fig:FC2} shows $E_d$ and $E_\phi$ as a function of $w$. As can be seen, formation control can be attained for $w>0$. However, our ability to control the network is lost as $w$ is increased above a certain threshold, $\omega \simeq 1.25$. This is in contrast with previous studies on formation control which assumed availability of exact measurements (see \eg Ref. \cite{ren2008distributed}) and suggests that the performance of the estimation strategy and control strategy are interdependent. The analysis of this interdependence, as well as of the stability of the desired formation configuration for the dynamics of the platforms and of the synchronous solution for the dynamics of the oscillators is the subject of ongoing investigations.
The performance of the control strategy may be improved by  a proper choice of the chaotic oscillators at the network nodes and of the network topology (\ie of the specific formation), as well as by the selection of the control strategy. This is also the subject of ongoing research.

\vspace{-4pt}
\section{Conclusions}\label{sec:Con}
In this paper we presented a decentralized framework to track variations of the network topology for a set of coupled mobile agents.
We proposed an adaptive technique based on synchronization of chaotic oscillators which, in the presence of limited information, was shown to be effective in synchronizing the oscillators and providing information on the local connectivity of the agents. For a network of three nodes,  each one of the agents can use the adaptive strategy to dynamically reconstruct the time evolution of the whole adjacency matrix and thus obtain  individual estimates of the distances to the other agents. We also found  the property of chaotic systems of being sensitively dependent on the initial conditions  convenient in terms of the estimation strategy (see Sec.\ \ref{Sec:Advantages}).

We further proposed an algorithm for formation control of a set of mobile agents moving in an unknown, uncertain, and time-varying environment, based on the distances estimated by the adaptive strategy. Our numerical simulations showed the effectiveness of this approach for a case in which estimation and control were simultaneously implemented. 

Future work will consist in extending our proposed approach to consider more complex networks with heterogeneous characteristics. By using subgroups of different chaotic oscillators, we could estimate the connections of large networks and discriminate between different categories or internal parameters for some of the agents in the system. We are also interested in investigating other control algorithms to avoid disconnections, while enforcing line-of-sight pathways among the robotic system.
Finally, we plan on testing the proposed strategy in an experiment with our test bed of ground and aerial vehicles \cite{bezzoTMECH}.

\section*{Acknowledgments}     
This work was supported in part by NSF grants ECCS \#1027775 and IIS \#0812338, and by the Army Research Laboratory grant \#W911NF-08-2-0004.

\end{document}